\documentclass[%
 amsmath,amssymb,
 aps, physrev,
10pt]{revtex4-2}
\usepackage{graphicx}
\usepackage{dcolumn}
\usepackage{bm}
\usepackage{iftex}
\ifPDFTeX
\else
  \errmessage{This document must be compiled with pdflatex!}
\fi

\usepackage{makecell}
\usepackage[linesnumbered,ruled,vlined]{algorithm2e}

\begin{document}

\preprint{} 

\title{\textbf{Clock Synchronization for Drone-Based Entanglement Quantum Key Distribution} 
}

\author{Jinquan Huang}
\affiliation{School of Electronics and Communication Engineering, Sun Yat-sen University, Shenzhen, 518107, Guangdong, China.}%
 \affiliation{College of Advanced Interdisciplinary Studies, National University of
Defense Technology, Changsha, 410073, Hunan,China.}%

\author{Bangying Tang}%
\affiliation{Strategic Assessments and Consultation Institute, Academy of Military Sciences, Beijing, China.}%

\author{JianJi Yi}%
\affiliation{School of Electronics and Communication Engineering, Sun Yat-sen University, Shenzhen, 518107, Guangdong, China.}%

\author{Bo Xu}%
\affiliation{School of Electronics and Communication Engineering, Sun Yat-sen University, Shenzhen, 518107, Guangdong, China.}%

\author{Hui Han}%
\affiliation{College of Advanced Interdisciplinary Studies,National University of
Defense Technology, Changsha, 410073, Hunan, China.}%

\author{Wanrong Yu}%
\affiliation{College of Advanced Interdisciplinary Studies,National University of
Defense Technology, Changsha, 410073, Hunan, China.}%

\author{Chunqing Wu}%
\affiliation{School of Electronics and Communication Engineering, Sun Yat-sen University, Shenzhen, 518107, Guangdong, China.}%

\author{Xiangwei Zhu}%
\affiliation{School of Electronics and Communication Engineering, Sun Yat-sen University, Shenzhen, 518107, Guangdong, China.}%

\author{Huicun Yu}%
\affiliation{Information and Navigation College, Air Force Engineering University,
Xi'an, 710077, Shaanxi, China.}%

\author{Jiahao Li}%
\affiliation{Information and Navigation College, Air Force Engineering University,
Xi'an, 710077, Shaanxi, China.}%

\author{Shihai Sun}%
\email{Contact author: sunshh8@mail.sysu.edu.cn}
\affiliation{School of Electronics and Communication Engineering, Sun Yat-sen University, Shenzhen, 518107, Guangdong, China.}%

\author{Bo Liu}%
\email{Contact author: liubo08@nudt.edu.cn}
 \affiliation{College of Advanced Interdisciplinary Studies, National University of
Defense Technology, Changsha, 410073, Hunan,China.}%


\begin{abstract}
Drone-based entanglement distribution provides full spatiotemporal coverage for quantum networks, enabling quantum key distribution (QKD) in dynamic environments. The security of QKD fundamentally depends on high-fidelity quantum state measurements, for which high-precision clock synchronization is indispensable, as timing jitter is inversely correlated with quantum state fidelity. However, drone-based clock synchronization is constrained by SWaP (Size, Weight, and Power) limitations and dynamic mobility effects. Here, we propose a synchronization protocol for drone-based entanglement distribution, leveraging nanosecond-accurate Global Navigation Satellite System (GNSS) timing and entanglement-based timing correction to overcome SWaP constraints. Experimental results demonstrate 24 ps RMS synchronization in simulated free-space quantum channels with distance dynamics, without requiring precision reference clock. Our protocol enables drone-based entanglement distribution, paving the way for seamless wide-area and local-area quantum internet.

\end{abstract}

\maketitle

\section{Introduction}
Entanglement-based QKD constitutes a foundational technology in quantum internet architectures. At present, entanglement-based QKD has been predominantly demonstrated through three principal implementations: fiber channel \cite{wengerowskyEntanglementDistribution96kmlong2019,wengerowskyPassivelyStableDistribution2020,neumannContinuousEntanglementDistribution2022,Liu_2024}  terrestrial free-space connections \cite{ursinEntanglementbasedQuantumCommunication2007,eckerStrategiesAchievingHigh2021,doi:10.1126/sciadv.abe6379,bassobassetDaylightEntanglementbasedQuantum2023,krzicMetropolitanFreespaceQuantum2023} and satellite to ground channels \cite{Yin_2017, yinEntanglementbasedSecureQuantum2020}. However, these platforms face inherent limitations in connection availability, fiber channel or terrestrial free-space channel due to fixed infrastructure deployment, and satellites due to orbital period constraints. Field demonstrations of drone-to-ground entanglement distribution have validated the prospect of drone-platforms application \cite{10.1093/nsr/nwz227, Liu_2021}. Drone platforms, leveraging their operational flexibility, maneuverability, and rapid deployment capabilities, are poised to enable seamless wide-area and local-area quantum networks. Extending this to drone-to-drone scenarios introduces new challenges for quantum state transmission, particularly in maintaining stable clock synchronization for coincidence measurements under SWaP and dynamic conditions.

The success of entanglement-based QKD hinges on the coincidence to accidental ratio photon-pair coincidence measurement, where precise inter-mode clock synchronization is critical. Since accidental coincidence rates scale with timing jitter, 
picosecond-level synchronization precision becomes essential to maintain the quantum state fidelity for secure key generation \cite{ neumannContinuousEntanglementDistribution2022}.  Recent advances in entanglement distribution employ several synchronization approaches: (i)  Global Navigation Satellite System (GNSS)\cite{eckerStrategiesAchievingHigh2021,doi:10.1126/sciadv.abe6379,bassobassetDaylightEntanglementbasedQuantum2023,Liu_2024},  (ii) Precision rubidium clock \cite{9720084, krzicMetropolitanFreespaceQuantum2023, peletEntanglementbasedClockSyntonization2025}, (iii) Clock synchronization using the hybrid method\cite{Yin_2017,wengerowskyEntanglementDistribution96kmlong2019,yinEntanglementbasedSecureQuantum2020}. Hybrid implementations combining these methods to adapt to actual deployment - for instance, GNSS and rubidium clock  exhibit 13ps/s clock drift \cite{neumannContinuousEntanglementDistribution2022}, while combined GNNS and synchronization pulses achieve sub-nanosecond jitter ($\approx 0.7$ ns) \cite{PhysRevLett.119.200501,yinEntanglementbasedSecureQuantum2020}. The White Rabbit protocol \cite{10589342}, providing picosecond-level clock synchronization, can also fulfill the synchronization requirements\cite{10727432, 7383303,Gerrits:22}. The entanglement source generates photon pairs with a coherence time of several hundred femtoseconds \cite{PhysRevLett.25.84}, which are subsequently detected by remote receivers and processed to produce coincidence peaks. The temporal shift of coincidence peaks encodes the relative propagation delay in the entangled photon link, enabling extraction of clock offsets between distant nodes \cite{10.1063/1.1797561, Ho_2009}. Entanglement-based quantum clock synchronization enables femtoseconds-level timing alignment across medium/long-haul nodes \cite{Shi:24} and multi-node network \cite{tangDemonstration75Kmfiber2023,liWavelengthMulticastingQuantum2024}, surpasses classical synchronization limits through quantum correlation. Alternatively, a classical reference-clock-assisted protocol first performs initial synchronization via coincidence measurements, then synchronization the clocks using temporal shifts of coincidence peaks-eliminating the need for bidirectional quantum state transfer. Through this approach, field-deployed quantum systems achieve synchronization precision ranging from tens to hundreds of picoseconds \cite{doi:10.1126/sciadv.abe6379,bassobassetDaylightEntanglementbasedQuantum2023,PhysRevX.13.021001,PhysRevApplied.19.054082,peletEntanglementbasedClockSyntonization2025}. However, drones' SWaP (Size, Weight, Power) constraints limit precision system deployment \cite{10.1093/nsr/nwz227, Liu_2021, PhysRevLett.133.200801}, while their mobility introduces time synchronization challenges that exacerbate clock uncertainty. This simplified precise clock synchronization architecture is suited for SWaP-constrained drone deployment. However, implementation on drone platforms remains challenging, particularly when using low-precision reference clocks and operating in quantum channels with dynamic distance variations.

In this paper, we propose a drone-based clock synchronization protocol for entanglement-based QKD. A significant innovation lies in this approach is to maintain precise clock synchronization under dynamic distances of quantum channel, allowing optimized QKD performance using low-precision and SWaP-constrained hardware. Our protocol suppress timing jitter through a two-stage process: (i) GNSS pulse per second (PPS) temporal alignment (PPS-TA), which extracts the mean timing offset of coincident photon pairs, followed by (ii) temporal-sliced entanglement correction (TSEC) to suppress residual timing jitter. We developed drone-mountable payload components and conducted laboratory emulation experiments that replicate the primary timing synchronization challenges encountered in practical deployments. For real-time operation, we implemented a dual-pointer fast correlation algorithm that completes corrections within milliseconds timescales. Our approach provides a pathway that harnesses quantum correlations and low precision reference clocks to overcome the major challenges in drone-based entanglement distribution, paving the way for full spatiotemporal coverage in future quantum internet infrastructure.

\begin{figure*}
  \includegraphics[width=1\linewidth]{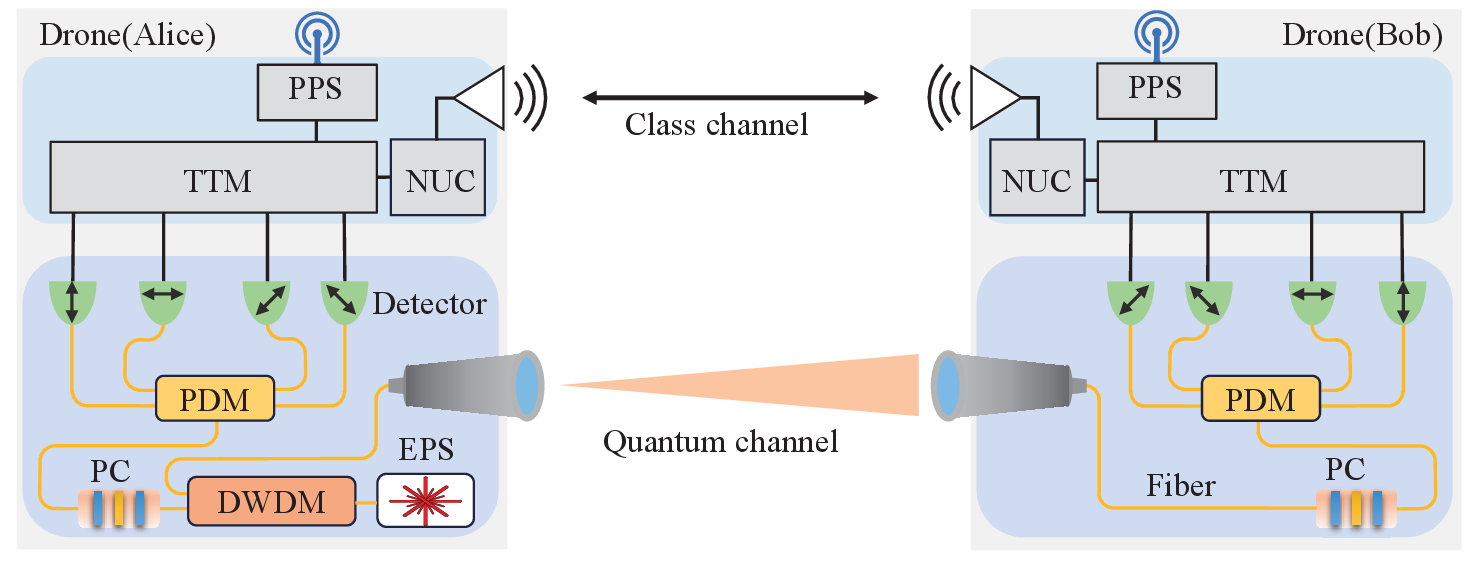}
  \caption{\label{fig:fig_1} 
  Drone-based entanglement-based QKD scheme.The polarization entangled photon source (EPS) is mounted on drone(Alice). A dense wavelength division multiplexer (DWDM) demultiplexes the broadband EPS spectrum. The signal photons are detected locally at Alice, while the idler photons are transmitted to Bob via free-space link. Both nodes employ identical photon detection and electronic systems. Photon are decoded by polarization modules (PDM) before detection. Time-tagging modules (TTM) record photons and PPS signal arrival times. Bob's timestamps form TTM are transmitted to Alice via classical channel for processing in the NUC(Next Unit of Computing). PC(Polarization Control).}
\end{figure*}

\section{\label{sec:II}Entangled Photon Pair Identification with drones}

Entangled photon pairs are detected through coincidence measurement, which record the arrival events both photons simultaneously, and identified by applying a coincidence window to filter. Fig.~\ref{fig:fig_1} illustrates a drone-to-drone entanglement-based quantum key distribution (QKD) implementation. Following propagation through lossy channels from the source, the photon pairs are measured by Alice and Bob, yielding quantum state click events $\{|\phi_i^A\rangle\}$ and $\{|\phi_j^B\rangle\}$, which are record as timestamp sequences $\left\{{t_i^A}\right\}$ and $\left\{{t_j^B}\right\}$. 
The identification of entangled photon pairs requires satisfaction of the following condition:
\begin{equation}
  \left\{
    \left\{ {\left[ {|\phi _i^A\rangle ,|\phi _j^B\rangle } \right]|\Delta t_{AB} - {\Delta t_\text{prop}}| \le {\tau _w}/2} \right\}
  \right\}.\label{eq:eq_1}
\end{equation}
where $\Delta t_{AB} = t_i^A - t_j^B$ represents the clock offset between Alice and Bob, and ${\Delta t_\text{prop}} = t_\text{prop}^A - t_\text{prop}^B$ accounts for the relative propagation delay between the communication parties.  ${\tau_w}$ denotes the coincidence window, typically chosen slightly larger than the timing jitter. Photon pairs satisfying  Eq.~(\ref{eq:eq_1}) are registered as coincidence events. Minimizing relative timing drift between Alice's and Bob's clock is critical for accurate identification coincidence events.  GNSS receivers, commonly used for drone positioning and attitude determination, provide external clock references through pulse-per-second (PPS) signals with typical nanosecond-level precision. Clock synchronization via PPS achieves initial accuracy of 10 ns $\sim$ 20 ns, with the ${\tau_w}$ being refined to sub-1.6 ns through coincidence events correction \cite{doi:10.1126/sciadv.abe6379,bassobassetDaylightEntanglementbasedQuantum2023}. The temporal correlations of entangled photon pairs enable precise synchronization of low-stability clock, achieving timing jitter $<$68 ps in fixed quantum channel \cite{PhysRevApplied.19.054082}. 

For drone-based deployment, we focus on implementing precise synchronization in dynamically varying quantum channels using SWaP-advantaged and low-precision devices. The GNSS-based synchronization scheme utilizes PPS signals as temporal references, aligning photon arrival times between consecutive pulses. However, nanosecond-scale jitter between pulses, combined with local clock drift and dynamic variations in relative propagation delays degrade temporal correlations. When integrated through intensity correlation functions, these factors broaden the coincidence peak. Real-time correction during coincidence measurement thus becomes essential for entangled pair identification with low-precision devices.

\section{\label{sec:III} Synchronization protocol}

This protocol sequentially reduces timing jitter through two stages: (i) PPS-based temporal alignment (PPS-TA) and (ii) temporal-sliced entanglement correction (TSEC). We characterize the synchronization performance boundaries under drone-based entanglement distribution scenario. Numerical simulations demonstrate the protocol's capability to maintain secure key rates within practical operational limits.
\subsection{\label{sec:subIII_A}Two-step synchronization architecture}
\begin{figure*}
  \includegraphics[width=0.8\linewidth]{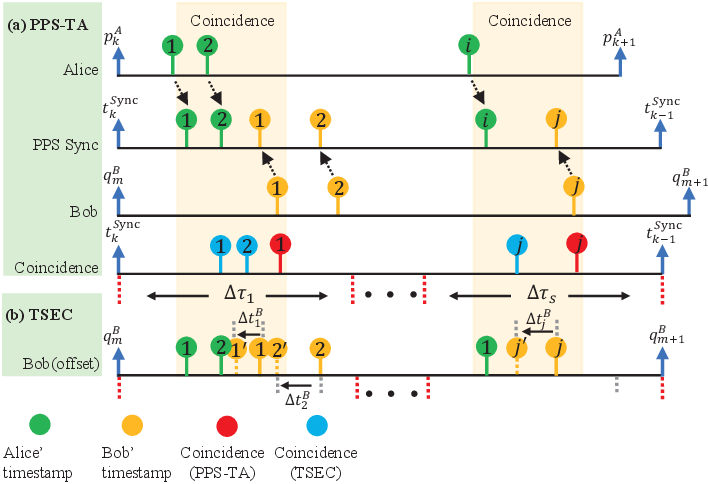}
  \caption{\label{fig:fig_2} Two-stage clock synchronization protocol for entanglement-based QKD. (a) First-stage coarse synchronization via PPS-TA, identifies coincidence events.(b) Second-stage TSEC precision synchronization employs entanglement correction with sliced su-blocks, utilizing entanglement correlations to suppress jitter and thereby increase detectable coincidence counts.}
\end{figure*}

 Our synchronization architecture employs a hierarchical approach combining PPS-TA with TSEC, as illustrated in Fig.\ref{fig:fig_2}. The first stage performs PPS-TA by coarse identification of coincidence events with PPS synchronization.  The second stage implements TSEC, where coincidence events are partitioned into temporal sub-blocks with for localized linearization of clock drift and propagation delay compensation.

\subsubsection{PPS-based Temporal Alignment}
In the GNSS-based clock synchronization system, GNSS time serves as the common reference. The PPS signal sequence for Alice and Bob are record as $\{p_k^{{\text{A}}}\}$ and $\{q_m^{{\text{B}}}\}$ by TTM. Accounting for jitter both the GNSS and TTM, the PPS signal can be expressed as: 
\begin{equation}
  p_m^{{\rm{A}}} = p_m^{{\rm{G}}} + \sigma_{{{p}}}^2\quad {\rm{and}}\quad q_m^{{\rm{B}}} = q_m^{{\rm{G}}} + \sigma_{{{q}}}^2.
\end{equation}
where $p_m^{{\rm{G}}} $ and $q_m^{{\rm{G}}}$ represent the standard GNSS time, and  $\sigma_{{{p}}}^2,\sigma_{{{q}}}^2 \sim{\cal N}(0,\sigma_{{\rm{PPS}}}^2) $. Alice and Bob alignment to a common clock reference using the PPS time sequence as their baseline: 
\begin{equation}
  t_i^{\text{A,Sync}} = \left(\frac{{t_i^\text{A} - p_{k}^\text{A}}}{{p_{k+n}^{\text{A}} - p_k^{\text{A}}}}\right)n,
  t_j^{\text{B,Sync}} =  \left(\frac{{t_j^\text{B} - q_{m}^\text{B}}}{{q_{m+n}^{\text{B}} - q_k^{\text{B}}}}\right)n.
\end{equation} 
where $n$ denotes the wide of the temporal alignment window, the residual clock can be given as:
\begin{equation}
\sigma_{{\rm{PPS-TA}}}^2 = \sigma_{{\rm{Sync,A}}}^2 + \sigma_{{\rm{Sync,B}}}^2 + \epsilon^2_\text{prop}
\end{equation} 
where $\epsilon^2_\text{prop}$ is denotes the rate of variation of relative propagation delay and 
\begin{equation}
  \sigma _{{\rm{A,Sync}}}^2 \approx \left[\frac{{\sigma _{{\rm{det}}}^2 + \sigma _{{\rm{PPS}},{\rm{A}}}^2}}{{{{\left( {p_{k + n}^{{\rm{A}}} - p_k^{{\rm{A}}}} \right)}^2}}} + \frac{{{{(t_i^{{\rm{A,Sync}}})}^2} \cdot 2\sigma _{{\rm{PPS}},{\rm{A}}}^2}}{{{{\left( {p_{k + n}^{{\rm{A}}} - p_k^{{\rm{A}}}} \right)}^2}}} \right]n, 
\end{equation} 
\begin{equation}
  \sigma _{{\rm{B,Sync}}}^2 \approx \left[\frac{{\sigma _{{\rm{det}}}^2 + \sigma _{{\rm{PPS}},{\rm{B}}}^2}}{{{{\left( {p_{k + n}^{{\rm{B}}} - p_k^{{\rm{B}}}} \right)}^2}}} + \frac{{{{(t_i^{{\rm{B,Sync}}})}^2} \cdot 2\sigma _{{\rm{PPS}},{\rm{B}}}^2}}{{{{\left( {p_{k + n}^{{\rm{B}}} - p_k^{{\rm{B}}}} \right)}^2}}}\right]n.
\end{equation} 
The synchronization error scales as $\sim1/n$ (see Appendix~\ref{Appendix:A} for details). While enlarging the window reduces synchronization error, this improvement need trade off real-time processing constraint in practical. The relative time clock error between Alice and Bob after PPS-TA processing is given as:
\begin{equation} 
\Delta t^{\rm{AB}} = t_i^{{\rm{Sync,A}}} - t_j^{{\rm{Sync,B}}} - \Delta {t_{{\rm{prop}}}} + \sigma_{{\rm{PPS-TA}}}^2
\end{equation} 
The timestamps sequences $\{t_i^{\text{A,Sync}}\}$ and $\{t_j^{\text{B,Sync}}\}$ are segmented and analyzed using intensity correlation functions to determine the mean temporal offset 
$\bar{\mu}$, which incorporates the relative propagation delay of photon pairs. The temporal broadening of the coincidence peak (manifested as increased timing jitter) stems from the cumulative integration of both PPS-TA timing jitter and equivalent delay variations induced by free-space channel fluctuations.
\subsubsection{Temporal-Sliced Entanglement Correction}

TSEC statistically analyzes entangled photon-pairs correlation within $ \tau_w$ to compensate time jitter in coincidence peak. The intensity correlation data blocks are subdivided into sub-blocks $\mathbf{T} = [\Delta {\tau_1},\Delta {\tau _2},...,\Delta {\tau_s}]$ for processing. $\Delta t $ can be approximated as linear within a relative small sub-intervals. Therefore, the offset can be linearly compensated. The mean offset of the sub-block can be expressed as $\mathbf{M} = [{\bar{\mu}_1},{\bar{\mu}_2},...,{\bar{\mu}_s}]$, where $\bar{\mu}_s = \frac{1}{{{N_s}}}\sum\limits_{k \in {\Delta {\tau_s}}} {\Delta t_k^{{\rm{AB}}}} $, with $N_s$ is the identified pairs of entangled photon within $\Delta {\tau_s}$, and the variance is given $\sigma_{{\mu_s}}^2 = \frac{{\sigma _{\Delta {t^{{\rm{AB}}}}}^2}}{{{N_s}}} = \frac{{\sigma_{{\rm{PPS-TA}}}^2}}{{{N_s}}}$. The clock at each timestamp within a sub-block can be approximated through linear interpolation (see Appendix~\ref{Appendix:B} for details):
\begin{eqnarray}
  \Delta t_j^{\rm{B}} =&& t_j^{{\rm{Sync,B}}} + \Delta{t_{{\rm{prop}}}} + \sigma_{{\rm{TSEC}}}^2
  \nonumber\\
  =&& \alpha({\mu_s}-{\mu_{s - 1}}) + {\mu_n} + \sigma_{{\rm{TSEC}}}^2
\end{eqnarray}
where the linear scaling $\alpha  = \frac{{t_j^B - {x_s}}}{{{x_{s}} - {x_{s-1}}}},\alpha  \in [0,1)$ is defined as the normalized position of timestamps, and $x_s$, $x_{s-1}$ denote adjacent reference point position, with $\Delta {\tau_s} = {x_{s}} - {x_{s-1}}$. The residual clock of TSEC is quantified by the variance:
\begin{equation} 
  \sigma_{{\rm{TSEC}}}^2 = \sigma_{{\rm{TS}}}^2 + \sigma_\mu ^2
\end{equation} 
where $\sigma_{{\rm{TS}}}$ is the timestamp jitter transfer error:
\begin{equation} 
\sigma _{{\rm{TS}}}^2 = \left| {\frac{{\partial (\Delta t_j^B)}}{{\partial t_j^B}}} \right|{\sigma _{t_j^B}} \approx (\frac{{{\sigma _{t_i^B}}}}{{\Delta {\tau _s}}})^2
\label{eq:10}
\end{equation} 
The Gaussian mean error $\sigma_\mu^2$ for sub-block $\mu_s$ undergoes proportional scaling through error propagation:
\begin{eqnarray}
\sigma _\mu ^2 =&& {\left( {\frac{{\partial (\Delta t_i^B)}}{{\partial {\mu _{s-1}}}}{\sigma _{{\mu _{s-1}}}}} \right)^2} + {\left( {\frac{{\partial (\Delta t_i^B)}}{{\partial {\mu _{s}}}}{\sigma _{{\mu _{s}}}}} \right)^2}
\nonumber\\
=&& {(1 - \alpha )^2}\sigma _{{\mu _{s-1}}}^2 + {\alpha ^2}\sigma _{{\mu _{s}}}^2
\label{eq:11}
\end{eqnarray}

\subsection{\label{sec:subII_B}Synchronization performance boundaries}
\subsubsection{Minimum temporal segmentation}
The duration of sub-block governs the validity of linear approximation for clock drift within each. $S_{\text{min}}$ can be determined based on $\sigma_\text{TSEC}$ within each integration period $T_s$. The segmentation criterion requires that the maximum second-order error induced by clock drift acceleration must not exceed the permitted linearization tolerance $\delta_\eta$, yielding(see Appendix~\ref{Appendix:C} for details):
\begin{equation} 
  S_{\text{min}} \geq T_n \sqrt{|\gamma|/(8\delta_\eta)}
  \label{eq:12}
\end{equation} 
where $|\gamma|$ represents the magnitude of clock drift acceleration. This formulation guarantees that the linear approximation remains valid across each sub-segment while maintaining the target timing precision. 
\subsubsection{Minimum photon-pairs per temporal segment}
The statistical significance of mean offset within each sub-block constitutes the critical determinant for the precision of linear compensation. Under the assumption of statistically equivalent adjacent sub-blocks with equal photon-pair counts($N_{s-1}=N_s$) and identical offset variances ($\sigma_{u_{s-1}}^2 = \sigma_{u_{s}}^2$). When the condition $\sigma_{{\rm{TSEC}}}^2 \le \sigma_{{\rm{PPS-TA}}}^2$ holds, the minimum bound for $N_s$ is derived as:
\begin{equation} 
{N_{\rm{s}}} \ge \frac{{\sigma _{{\rm{PPS-TA}}}^2}}{{2\left[ {\sigma _{{\rm{PPS-TA}}}^2 - {{\left( {\frac{{{\sigma _{t_i^B}}}}{{\Delta {\tau _s}}}} \right)}^2}} \right]}}
\end{equation} 
In practical implementation, the condition ${T_s} \gg {\sigma _{t_i^B}}$, ensures $N_s \ge 0.5$. This indicates that when a single photon pair is detected within $\Delta \tau_s$, the achieved timing precise surpasses the $\sigma_\text{PPS-TA}$ baseline. Furthermore, we must account for the Gaussian mean significance of $\mu_s$. To satisfy a given statistical error bound ${\sigma _\mu }\le \varepsilon $, the minimum entangled photon pairs requirements given as (see Appendix~\ref{Appendix:D} for details): 
\begin{equation}
{N_s} \ge {\left( {\frac{{{\sigma _t}}}{\varepsilon}} \right)^2}
\end{equation} 
These dual constraints establish a foundational trade-off in sub-blocks partitioning. While increasing the number of sub-blocks reduces linear fitting error and thereby improves compensation accuracy, it simultaneously decreases the number of entangled of entangled photon pairs per sub-block. This reduction in pair density degrades the statistical significance of mean offset estimation, ultimately compromising the compensation precision.
\subsection{\label{sec:subIII_C}Simulation analysis}

 \begin{figure*}
  \includegraphics[width=1\linewidth]{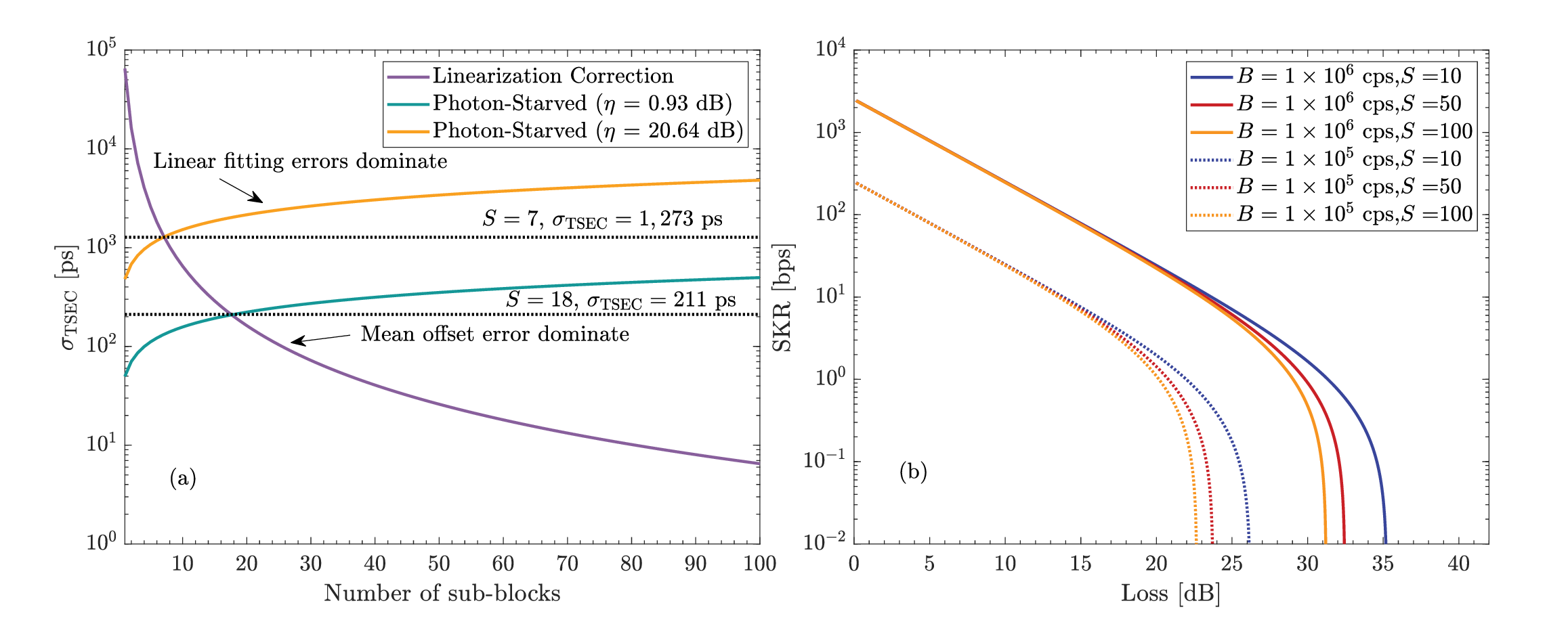}
  \caption{\label{fig:fig_3}  Simulation of clock synchronization performance. (a) Trade-offs between sub-block count, channel loss, and timing jitter. (b) Dependence of secure key rate on sub-block partitioning schemes (with corresponding residual timing jitter).} 
\end{figure*}
Based on the clock synchronization model and entanglement distribution model (detailed in Appendix~\ref{Appendix:E}), which parameter from Tables~\ref{tab:table1}, we evaluated timing jitter characteristics in a drone-based QKD scenario.

Fig.~\ref{fig:fig_3}(a) demonstrates the effect of sub-block partitioning on $\sigma_{\text{TSEC}}$. Using a fixed data block period ($T_s = 1$ s) and drift acceleration ($\gamma = 0.3 \text{ ps/ms}^2$, equivalent to 0.1 m/s channel distance variation), we first establish the minimum sub-block count required to achieve target $\sigma_{\text{TSEC}}$ values. Increasing sub-block counts in the linearized correction regime reduces jitter through finer temporal resolution and decreased linear fitting errors.  Conversely, in the photon-starved regime with fixed source brightness at 0.93 dB (200 m) and 20.64 dB (1000 m) channel losses, additional sub-blocks reduce entangled photon counts per sub-block, thereby increasing mean offset errors. Through this trade-off analysis, we identify the optimal sub-block count for practical implementation. 

For a given brightness $B$, we observe an inverse relationship between number of su-block $S$ and SKR $R$, where fewer sub-blocks yield higher SKR due to reduced mean offset errors in synchronization, as illustrated in Fig.~\ref{fig:fig_3}(b). This phenomenon originates from three competing factors:(i) the statistical advantage of larger sub-blocks (smaller $S$) providing increased entangled photon pairs per block ($N_s \propto 1/S$), which improves the signal-to-noise ratio in offset estimation ($\sigma_\mu \propto 1/N_s $); (ii) the scaling of timestamp jitter with sub-block duration ($\Delta \tau_s/S$) as derived in Eq.\ref{eq:10}, where coarser partitioning limits error propagation; and (iii) the nonlinear drift dominance when $S$ falls below the $S_\text{min}$ threshold in Eq.\ref{eq:12}, manifesting as characteristics in infection point in Fig.~\ref{fig:fig_3}(a).

\begin{table}[b]
\caption{\label{tab:table1}
Simulation parameter of drone-based entanglement QKD.}
\begin{ruledtabular}
\begin{tabular}{cccccc}
 &Value& Unit & & Value & Unit \\
\hline
B& $1 \times {10^6}$ & cps & DT & 24.6 & mm \\
$\sigma_c$& 3 & ps &DR & 24.6 & mm \\
$\sigma_\text{Det}$& 45 & ps &$\theta_\text{E}$ & 0 & rad \\
$\sigma_\text{TTM}$& 45 & ps &$\theta_\text{E}$ & 0 & rad \\
$\sigma_\text{p}$& 1000 & ps & $\lambda$ & 1550 & nm\\
$f$ & 1.09 & - & $r_0$& 0.2 & - \\ 
$e_0$ & 0.01 - & - & $\tau_0$ & 0.02 & - \\
$\text{DC}_\text{A}$ &  1000 & cps & $\sigma_T$ &  $6 \times {10^6}$ & -\\
$\text{DC}_\text{A}$& 1000 & cps &  $n_0$ & 1.0 & - \\
 &  & & SR & 1.0 & -\\
\end{tabular}
\end{ruledtabular}

\end{table}

\section{\label{sec:IV}Tabletop Experiments}
\begin{figure*}
  \includegraphics[width=1\linewidth]{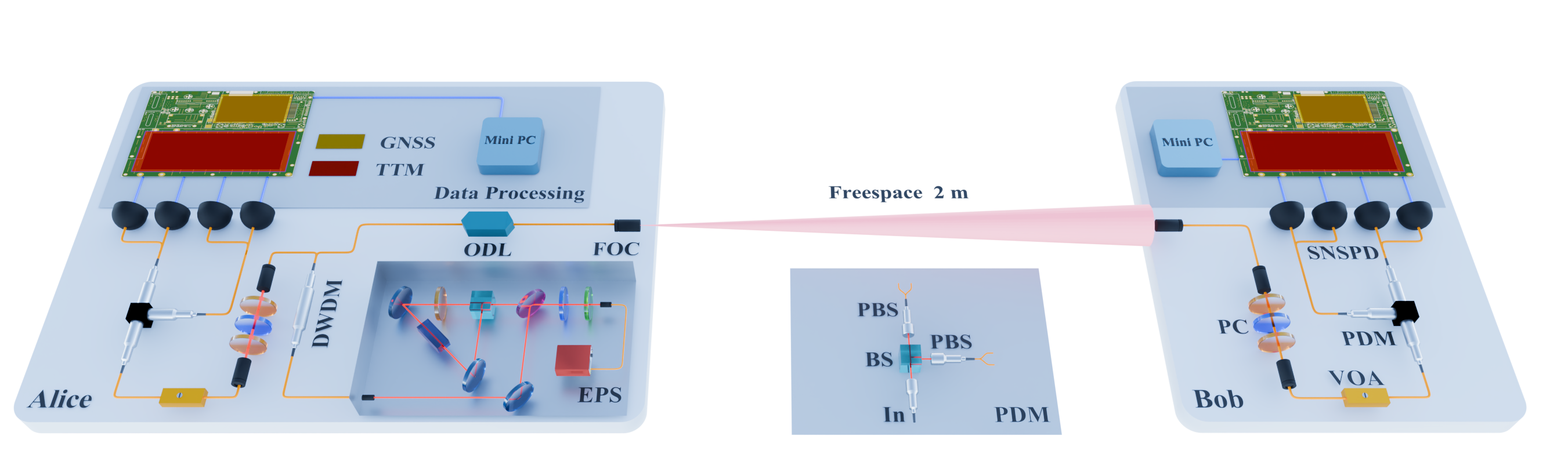}
  \caption{\label{fig:fig_4} The tabletop experimental setup for drone-based entanglement-based QKD. The free-space quantum channel is experimentally realized in a laboratory environment, where an ODL (optical delay line) emulates practical relative delay jitter and  VOA (variable optical attenuator) control link loss. The integrated polarization decoded module (PMD) measurement quantum states in a set of mutually unbiased bases. Data processing with TTM recording photon arrival times and PPS signal, processed by a Mini-PC} 
\end{figure*}

The experiment physically replicates drone entanglement-based QKD quantum channel variation. Fig.~\ref{fig:fig_4} illustrates the experimental setup. A polarization source generating photon pairs in the Bell state \(|\psi^+\rangle = \frac{1}{\sqrt{2}}(|H\rangle_A |H\rangle_B + |V\rangle_A |V\rangle_B)\), with wavelength channels C33 (to Alice) and C35 (to Bob) allocated via DWDM. The free space channel incorporates: (i) optical delay lines simulating 300 ps/s relative temporal variation (equivalent to 0.1 m/s drone motion), and (ii) variable optical attenuator emulating channel losses. Polarization drift is compensated before measurement using a polarization decoding  module (90\% efficiency) and superconducting nanowire detectors  (80\% efficiency). Photon arrival times and GNSS PPS (jitter with 10 ns $\sim$ 20 ns) are recorded using a time-tagging module (TTM), while timestamps from Bob are transmitted to Alice via a classical communication channel for coincidence processing. The TTM exhibits 45 ps RMS jitter, with its interval 10 MHz crystal oscillator demonstrates frequency stability of $\pm1.0$ppm/year.

\begin{figure*}
  \includegraphics[width=1\linewidth]{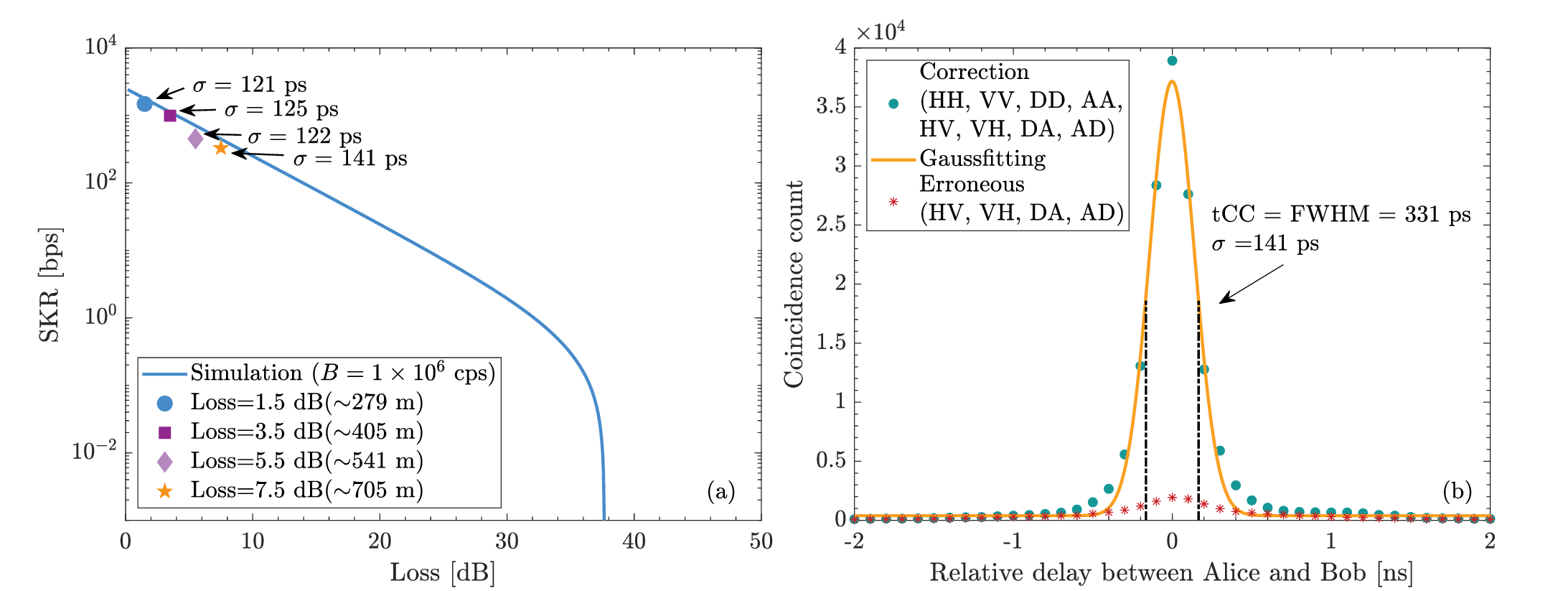}
  \caption{\label{fig:fig_5} (a) Secure key rate versus channel loss. Comparison of experimental results (independent markers) and numerical simulation (curves). The free-space distance is represented by equivalent channel loss. (b) Intensity correction function $g^{2}(\tau)$ for all coincidence used for sifted detection (basis matched coincidences), integrated over 100 s, with relative delay between Alice an Bob between detectors has been subtract. The coincidence window was configured to match the full width at half maximum (FWHM) of the Gaussian-fitted function.}
\end{figure*}
To evaluate the robustness of two-stage clock synchronization protocol under loss condition, we conducted a series of controlled tabletop experiments with varying channel losses. The losses are 1.5 dB, 3.5 dB, 5.5 dB and 7.5 dB respectively, and the equivalent distances are 279 m, 405 m, 541 m and 705 m, as shown in Fig.~\ref{fig:fig_5}(a). The SKR was recorded using the synchronization protocol, while numerical simulations modeled the variation in SKR at fixed brightness as a function of channel loss. To ensure robust clock synchronization across different loss condition, we setting the protocol with sub-block $S = 20$. Fig.~\ref{fig:fig_5}(b) presents the intensity correlation function for the 7.5 dB loss cases, revealing a characteristic of 135 ps. During a 100 s acquisition period, the system generated $94.91\times10^3$ bit coincidence events with $5.35\times10^3$ erroneous bits, yielding a quantum bit error rate (QBER) of 5.63\% and a security key rate (SKR) of 259 bps. Our results demonstrate that within the tested loss configurations, timing jitter showed no observable increase with rising channel attenuation.

\begin{figure}[b]
  \includegraphics[width=1\linewidth]{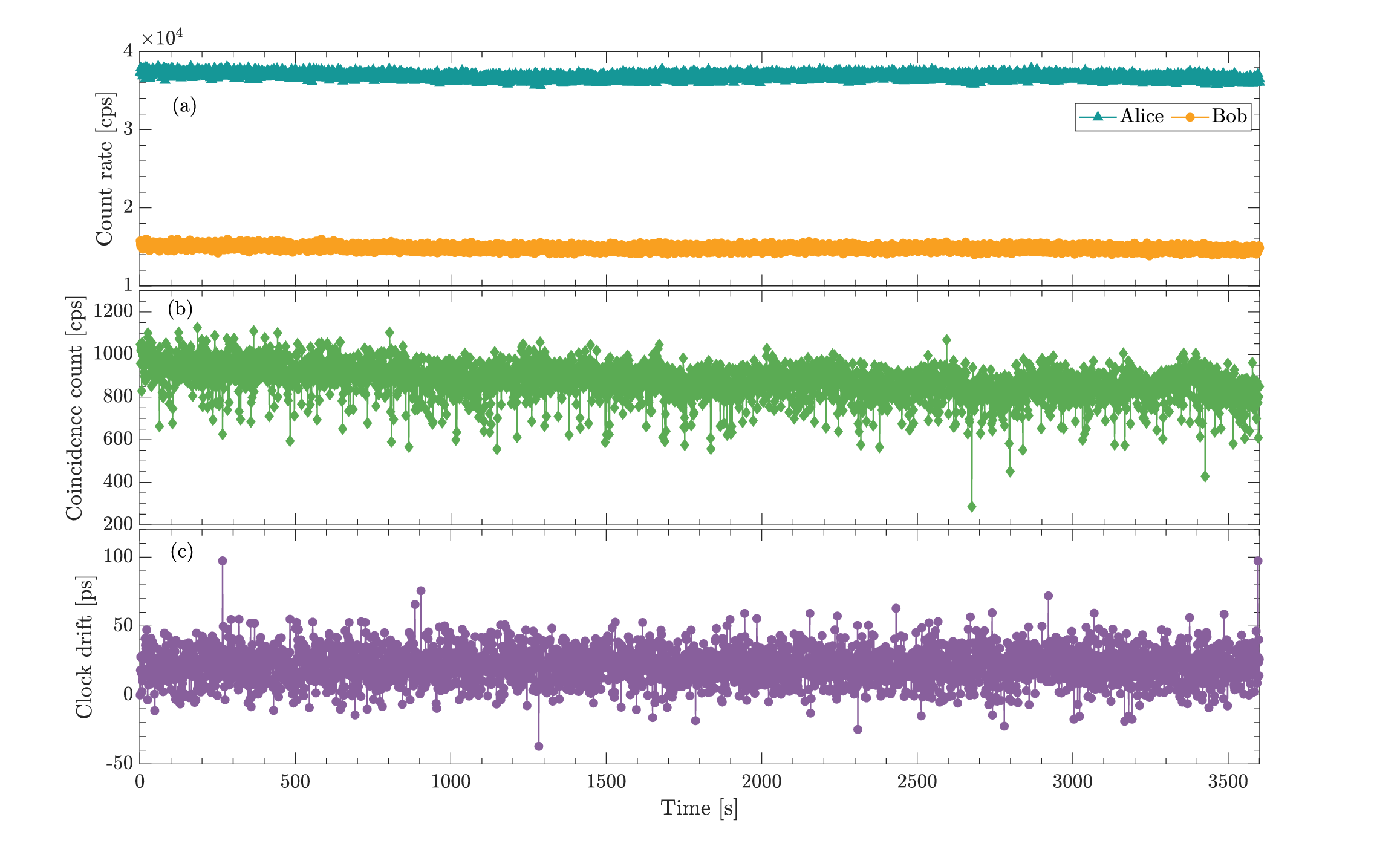}
  \caption{\label{fig:fig_6} One-hour continuous measurement of system stability. (a) Alice's and Bob's photon count rate as functions of time. (b) Coincidence count rate over time. (c) Clock drift (derived from coincidence peak offset) over time. }
\end{figure}

To validate the stability of our clock synchronization protocol, we conducted continuous 1-hour measurements under the 7.5 dB channel loss. Fig.~\ref{fig:fig_6} (a) photon count rates showing Alice's detector at $36.30\times10^3\pm234$ (root-mean-square $\pm$ standard deviation) cps and Bob's at $14.80\times10^3\pm355$ cps. The correction coincidence count rate is measured at 877$\pm$80 cps. Our clock synchronization protocol operates on each valid coincidence channel pair, demonstrating stable performance with a average coincidence count rate 219 cps per channel pair. For linear drift compensation, we employed a coincidence window slightly wider than the FWHM to ensure capture of correlated photon pairs. Fig.~\ref{fig:fig_6}(b) characteristic the clock drift around 3600 s. We performed outlier detection using the  median absolute deviation (MAD) method with threshold of $>10\times$MAD, then eliminating 102 anomalous data points (2.83\% of total samples) , retaining 97.17\% valid measurements. The  clock drift demonstrates stability 24$\pm$12 ps.

\section{\label{sec:V}Discussion}
High-precision clock synchronization constitutes a fundamental requirement for advanced entanglement distribution, as demonstrated in Appendix~\ref{Appendix:G}. Our protocol achieves 24$\pm$12 ps RMS clock drift, comparable to state-of-the-art systems employing synchronization pulses or rubidium clocks (Detailed in Appendix \ref{Appendix:D} and Table~\ref{tab:qkd_comparison}), while operating with low stability GNSS PPS and oscillators through our two-stage PPS-TA and TSEC jitter suppression technique. The TTM employs a crystal oscillator exhibiting $\pm1.0$ ppm/year frequency stability - substantially inferior to the $<0.5\times10^{-3}$ ppm/year stability of rubidium clock (Rb-clock) used in long-distance entanglement-based QKD, which exhibits a relative clock drift 13 ps/s when GNSS disciplined \cite{neumannContinuousEntanglementDistribution2022,FS740}. The system maintains tens of picoseconds precision in free-space quantum channels with 300 ps/s delay variations, outperforming GNSS-based solutions that exhibit hundreds of picoseconds jitter in static free-space configurations \cite{doi:10.1126/sciadv.abe6379,bassobassetDaylightEntanglementbasedQuantum2023}. Notably, the protocol sustains stable operation at coincidence rates as low as 219 cps per channel pair without introducing additional link loss \cite{PhysRevX.13.021001}. Furthermore, the demonstrated robustness against dynamic delays suggests potential applicability to long-distance fiber where comparable timing variations occur \cite{wengerowskyPassivelyStableDistribution2020}.

Our work focuses on drone deployment with rigorous SWaP optimization for constraint. The integrated design combines TDC and GNSS receivers on a 20 cm $\times$ 20 cm printed circuit board weighing 0.3 kg. The implementation of the two-stage synchronization protocol relies on three calculations of the intensity autocorrelation function. We have implemented a two-pointer intensity autocorrelation algorithm with O(n) complexity that process correlation measurement of $144\times10^3$ (Alice) and $133\times10^3$ (Bob) detection events in $7.49\pm0.74$ ms on a low-performance mini-PC (Detailed in Appendix \ref{Appendix:D}). By comparison, existing implementations with similar data volumes typically require seconds to tens of seconds for correlation calculations \cite{PhysRevApplied.19.054082,peletEntanglementbasedClockSyntonization2025}. The optical subsystem incorporates a polarization entanglement source (12 cm $\times$ 20 cm) maintaining $>$99\% fidelity, and PDM with the size of 40 mm $\times$ 70 mm with the efficiencies $>90\%$. With the commercial drones typically offering 30 kg $\sim$ 40 kg payload capacity \cite{dji}, our SWaP-optimized design demonstrates field deployment compatibility. We employ SNSPDs with high efficiency and low jitter, noting recent airborne SNSPDs achieve more than $90\%$ detection efficiency - confirming their feasibility for onboard implementation \cite{10.1117/1.APN.4.2.026003}. This SWaP optimization enables practical entanglement-based QKD in mobile field deployments.

Our results establish a foundation for exploring the potential of drone-based entangled quantum key distribution, paving the way for both fundamental quantum science investigations and practical technological applications in this emerging field.

\section{\label{sec:VI}Conclusion}
In this work, we have experimentally demonstrated a robust clock synchronization protocol for drone-based entanglement QKD.Our two-stage synchronization architecture combines PPS time alignment with time-sliced entanglement correction while establishing fundamental bounds on entanglement resource requirements. The protocol achieves 24$\pm$12 ps RMS synchronization precision under varying relative delay free-space channel condition. These results advance drone-based clock synchronization for entanglement QKD, particularly for dynamic scenarios demanding precise timing amid channel variations.

\begin{acknowledgments}
This work was supported by the Shenzhen Science and Technology Program \\ (JCYJ20220818102014029) and the Science and Technology Innovation Program of Hunan Province (2023RC3003). We sincerely appreciate their generous financial support.
\end{acknowledgments}

\appendix

\section{\label{Appendix:A}PPS-TA error analysis}
We analyze the synchronization error in PPS-TA using error propagation theory \cite{bouleauHistoricalAspectsError2015}. For a fractional $z = \frac{x}{y}$, the variance propagates as:
\begin{equation}
  \sigma _z^2 \approx {\left( {\frac{{\partial z}}{{\partial x}}} \right)^2}\sigma _x^2 + {\left( {\frac{{\partial z}}{{\partial y}}} \right)^2}\sigma _y^2.
\end{equation}
Applying this to Alice's timestamps:
\begin{equation}
  t_i^{\text{A,Sync}} = \left[\frac{{t_i^\text{A} - p_{k}^\text{A}}}{{p_{k+n}^{\text{A}} - p_k^{\text{A}}}}\right]n
  =\frac{X}{Y}n
\end{equation}
where the numbering variance is $\sigma _X^2 = \sigma _{{\rm{det}}}^2 + \sigma _{{\rm{PPS-TA}}}^2$. Assuming statistical independent between $t_{p + n}^{{\rm{PPS,A}}}$ and $t_{p}^{{\rm{PPS,A}}}$, the denominator variance is $\sigma_{Y}^2 = 2 \sigma_{\rm{PPS-TA}}^2$. 
Since  $\sigma_{\text{det}}^2$ and $\sigma_p^2$ are uncorrelated,$\text{Cov}(X,Y) = 0$. The synchronized timestamp variance thus propagates 
\begin{equation}
\sigma _{{\rm{Sync}},{\rm{A}}}^2 \approx \left[{\left( {\frac{1}{Y}} \right)^2}\sigma _X^2 + {\left( {\frac{X}{{{Y^2}}}} \right)^2}\sigma _Y^2\right]n.
\end{equation}
Substituting $X = t_i^{{\rm{Sync,A}}} \cdot Y$ yields
\begin{equation}
\sigma _{{\rm{Sync}},{\rm{A}}}^2 \approx \left[\frac{{\sigma _{{\rm{det}}}^2 + \sigma _{{\rm{PPS}},{\rm{A}}}^2}}{{{{\left({p_{k+n}^{\text{A}} - p_k^{\text{A}}} \right)}^2}}} + \frac{{{{(t_i^{{\rm{Sync}},{\rm{A}}})}^2} \cdot 2\sigma _{{\rm{PPS}},{\rm{A}}}^2}}{{{{\left({p_{k+n}^{\text{A}} - p_k^{\text{A}}} \right)}^2}}}\right]n.
\end{equation}
The contributions of the term $t_i^{{\rm{Sync}},{\rm{A}}} \subset [0,1]$  to the total error is determined by its temporal position within the sub-block. Crucially, all error scale as $1/n$, where $n = [{p_{k+n}^{\text{A}} - p_k^{\text{A}}}]$ and $[\cdot]$ denotes rounding to the nearest integer.

\section{\label{Appendix:B}TSEC error analysis}

Assuming adjacent sub-blocks contain similar numbers of entangled photon pairs $N_s \approx N_{s-1}$ and statistical characteristics 
$\sigma_{\mu_s} \approx \sigma_{\mu_{s-1}} $.The Eq.~(\ref{eq:11}) can be simplified to:
\begin{equation}
\sigma_u^2 = \left[ {{{(1 - \alpha )}^2} + {\alpha ^2}} \right]\sigma_{\mu_s} ^2
\end{equation}
Through error propagation analysis with $\frac{{d {\sigma_u^2}}}{{d\alpha }} = 0$  and $\frac{{d {\sigma_u^2}}}{{d\alpha^2}}>0$, we identify $\alpha = 0.5$ as the minimization point, representing the most balanced error weighting that yields the maximum absolute error. When $\alpha \rightarrow 0$ or $\alpha \rightarrow 1$ , the synchronization depends predominantly on a single sub-block mean. For asymmetric cases where ${N_{s-1}} \ll {N_{s}}$, the worst-case scenario occurs when $ \alpha \rightarrow 0 $, as this maximizes the contribution form the $\sigma_{u_{s-1}}$

  \section{\label{Appendix:C} Cramér-Rao bound on entangled photon pairs of sub-block}

    Consider the behavior of the clock deviation function $ \Delta t(t) $ over the time interval $T_n$. For any sub-segment $ \Delta \tau_s = \left[ {t_0, t_0 + \Delta \tau_s} \right] $, a Taylor expansion around the midpoint $ t_C = t_0 + \Delta \tau_s / 2 $ yields:

  \begin{equation}
    \Delta t(t) \approx \Delta t(t_C) + \eta(t_C)(t_0 - t_C) + \frac{1}{2}\gamma(t_C)(t_0 - t_C)^2
  \end{equation}
  where $\eta(t_C)$ is the first derivative (drift velocity) and $\gamma(t_C)$ is the second derivative (drift acceleration). When fitting a linear model $\Delta \hat{t}(t) = a(t_0 - t_C) + b$, the maximum error occurs at the boundaries of the sub-segment:
  \begin{equation}
    |\Delta t(t) - \Delta \hat{t}(t)| \leq \frac{1}{2}|\gamma(t_C)|(t - t_C)^2 = \frac{|\gamma| \Delta \tau_s^2}{8}.
  \end{equation}
  To ensure the validity of the linear approximation, the following condition must be satisfied:

\begin{equation}
  \frac{|\gamma| \Delta \tau_s^2}{8} \leq \delta_\eta,
\end{equation}
where $\delta_\eta$ is the maximum allowable linear error tolerance, related to the final target accuracy $\sigma_{\text{TSEC}}$ by $3\sigma_{\text{noise}} < \delta_\eta \approx \frac{\sigma_{\text{TAR}}}{\sqrt{3}}$ (If \(\delta_\eta\) is smaller than the noise level, it may lead to excessive segmentation.). The linear error is thus tied to the target accuracy (final precision). Consequently, the minimum number of sub-segments required is: 
\begin{equation}
S_{\text{min}} \geq T_n \sqrt{\frac{|\gamma|}{8 \delta_\eta}}.
\end{equation}

Example: For a target accuracy of $\sigma_{\text{TSEC}} = 150$ ps, $\delta_\eta = 87$ ps, and a drift acceleration of $\gamma = 0.3 \text{ps}/\text{ms}^2$ , we obtain $S_{\text{min}} = 21.8$. 

\section{\label{Appendix:D} Cramér-Rao bound on entangled photon pairs of sub-block}

The minimum entangled photon-pair count per sub-block is determined through Cramér-Rao bound analysis and Neyman-Pearson criterion verification \cite{Cramér+1946,doi:10.1098/rsta.1933.0009}. Consider the timing offset vector a sub-block with $N$ photon pairs, ${\mathbf{N}} = [\Delta t_1^{{\rm{AB}}},\Delta t_2^{{\rm{AB}}}, \ldots ,\Delta t_N^{{\rm{AB}}}]$,where each offset follows $\Delta t_i^{{\rm{AB}}} = \tau _i^{{\rm{AB}}} + {\varepsilon _i}$, with $\tau _i^{{\rm{AB}}}$ being the true offset and ${\varepsilon _i} \sim {\cal N}(0,\sigma _{{\rm{TSEC}},i}^2)$. The sub-block mean offset $\mu$ is given by 
\begin{equation}
\mu  = \frac{1}{N}\sum\limits_{i = 1}^N {\Delta t_i^{{\rm{AB}}}}  = \frac{1}{{{N_n}}}\sum\limits_{i = 1}^{{N_n}} {\left( {\,\tau _i^{{\rm{AB}}} + {\varepsilon _i}} \right)} \end{equation}
The error term $\mu  - {\mu _\tau } = \frac{1}{{{N_n}}}\sum\limits_{i = 1}^{{N_n}} {{\varepsilon _i}}$ follows a Gaussian distribution, where ${\mu _\tau } = \frac{1}{N}\sum\limits_{i = 1}^N {\tau _i^{{\rm{AB}}}}$. ${\varepsilon _i}$ is an independent Gaussian random variable, and their linear combination remains Gaussian. Consequently, $ \mu  - {\mu _\tau }\sim{\cal N}\left( {0,\frac{1}{{{N^2}}}  \sum\limits_{i = 1}^N {\sigma {{_{{\rm{TSEC,}}}^2}_i}} } \right)$. Moreover, each offset shares identical variance, this simplified to $\mu  - {\mu _\tau }\sim{\cal N}\left( {0,\frac{{\sigma _{{\rm{TSEC}}}^2}}{N}} \right)$.

The Cramér-Rao bound establishes the minimum variance achievable by any unbiased estimator. For our estimator $\mu$ of $\mu_\tau$, we derive the lower bound through Fisher information analysis.The likelihood function for each $\Delta t_i^{{\rm{AB}}}$ is given by:

\begin{equation}
p(\Delta t_i^{{\rm{AB}}}|\tau _i^{{\rm{AB}}}) = \frac{1}{{\sqrt {2\pi \sigma {{_{{\rm{TSEC}}}^2}_i}} }}\exp \left( { - \frac{{{{(\Delta t_i^{{\rm{AB}}} - \tau _i^{{\rm{AB}}})}^2}}}{{2\sigma {{_{{\rm{TSEC}}}^2}_i}}}} \right)
\end{equation}

Under the i.i.d. assumption, the joint likelihood function becomes $p({\mathbf{N}}|{{\mathbf{\tau }}^{{\rm{AB}}}}) = \prod\limits_{i = 1}^N p (\Delta t_i^{{\rm{AB}}}|\tau _i^{{\rm{AB}}})$, with corresponding log-likelihood: 
\begin{equation}
\log p({\bf{N}}|{{\bf{\tau }}^{{\rm{AB}}}}) =  - \frac{1}{2}\sum\limits_{i = 1}^N {\left[ {\log (2\pi \sigma _{{\rm{TSEC,}}i}^2) + \frac{{{{(\Delta t_i^{{\rm{AB}}} - \tau _i^{{\rm{AB}}})}^2}}}{{\sigma _{{\rm{TSEC,}}i}^2}}} \right]}
\end{equation}
The Fisher information for $\mu_\tau$ is: 
\begin{equation}
I({\mu _\tau }) =  - \left[ {\frac{{{\partial ^2}}}{{\partial \mu _\tau ^2}}\log p({\mathbf{N}}|{{\mathbf{\tau }}^{{\rm{AB}}}})} \right] =  - \sum\limits_{i = 1}^N {\frac{1}{{\sigma _{{\rm{TSEC,}}i}^2}}}
\end{equation}
Assuming identical true arrival times ${\mu _\tau } = \tau _i^{{\rm{AB}}}$ (i.e., identical true arrival times), the variance lower bound simplified to :
\begin{equation}
{\rm{Var}}({\hat \mu _\tau }) \ge \frac{{\sigma _{{\rm{TSEC}}}^2}}{N}
\end{equation}
We validate this bound via Neyman-Pearson hypothesis testing at 95\% confidence coefficient: 
(a) ${H_0}:u = {\mu _\tau }$(null hypothesis)
(b) ${H_1}:{H_0}:u \ne {\mu _\tau }$(alternative hypothesis).
The test statistic follows $Z = \frac{{\mu - {\mu _\tau }}}{{\sigma _{{\rm{TSEC}}}^2/\sqrt N }}\sim{\cal N}(0,1)$,with the reject region ${\rm{|Z| > 1}}{\rm{.96}}$. We obtain the minimum photon-pair requirements:
\begin{equation}
{N_n} \ge {\left( {\frac{{1.96{\sigma _{{\rm{TSEC}}}}}}{\gamma }} \right)^2}
\end{equation}
where the $\gamma$ represents the maximum tolerable error. For typical parameter ${\sigma _{{\rm{TSEC}}}} = 100$ ps and  $\gamma  = 50$ ps,$N \ge 15.36$.

\section{\label{Appendix:E} Simulated security key rate generation}
Our simulation of secure key generation is based on a continuous-wave entangled photon distribution model that incorporates dominant contributions to the QBER: system timing uncertainty (which reduces photon-pairs correlation), free-space transmission losses. 
\subsection{\label{app:subsec_F1}System time uncertainty}
The total time uncertainty experienced by the drone quantum entanglement distribution of entangled photon pairs is
  \begin{equation}
  \Delta T = \sqrt {\sigma _J^2 + \sigma _T^2}
 \end{equation}
 Here,  $\sigma _J^2 = \sqrt{\sigma _C^2 + \sigma_\text{Det} + \sigma_\text{TTM}}$ denotes the total system jitter, comprising contributions from the photon coherence time $\sigma _C^2$, detector timing jitter $\sigma_\text{Det}$, and TTM jitter $\sigma_\text{TTM}$. The additional term $\sigma _T^2$ accounts for residual errors in time synchronization.
\subsection{\label{app:subsec_F2}Free-space channel loss}
 The spatial intensity distribution of the optical beam between drone nodes follows a Gaussian profile:
\begin{equation}
I(r,l) = {I_0} \times {e^{\frac{{ - 2{r^2}}}{{\omega _0^2}}}}
\end{equation}
where ${I_0} = \frac{{2P}}{{\pi \omega _0^2}}$ is the peak intensity derived from total power $P = \int_0^\infty  {I\left( r \right)}  \cdot 2\pi r\;dx = {I_0}\frac{{\pi \omega _0^2}}{2}$, with $\omega _0$ representing the effective beam waist at the receiver:
\begin{equation}
{\omega _0} = \sqrt {\omega _L^2 + {{\left( {{\sigma _T} \cdot l} \right)}^2}}
\end{equation}
where ${\sigma _T}$ demotes transmitter pointing error, while ${\omega _L}$ accounts for diffraction-limited beam spreading:
\begin{equation}
{\omega _L} = l\frac{\lambda }{{\pi  \cdot 0.316{D_T}}}{\left[ {1 + 0.83 \cdot \sec (\theta ){{\left( {\frac{{{D_T}}}{{{r_0}}}} \right)}^{5/3}}} \right]^{3/5}}
\end{equation}
The efficiency of link is given as:
\begin{equation}
{\eta _0} = {\rm{SR}} \cdot \exp [ - \tau  \cdot \sec ({\theta _E})] \cdot \left\{ {1 - \exp \left[ { - 0.5{{\left( {\frac{{{D_R}}}{{{\omega _{L{\rm{P}}}}}}} \right)}^2}} \right]} \right\}
\end{equation}
where ${D_R}$ denotes the effective aperture diameter of the receiving telescope, $\tau $ represents the atmospheric optical depth at the operating wavelength, and ${\theta _E}$ is the elevation angle of the telescope relative to horizon.
where the Strehl ratio (SR) characterizes wavefront distortion: 
\begin{equation}
{\rm{SR}} = \frac{{{I_r}}}{{{I_0}}} \approx \exp \left( { - \phi _{{\rm{rms}}}^2} \right) = \exp \left[ { - {{\left( {\frac{{2\pi OP{D_{{\rm{rms}}}}}}{\lambda }} \right)}^2}} \right]
\end{equation}
 \subsection{\label{app:subsec_F3}Security key rate}
 The total measurement error of entanglement distribution can be expressed as \cite{neumannModelOptimizingQuantum2021,Neumann_2021}
 \begin{equation}
 E = \frac{{{\eta ^\tau }B{\eta _{\rm{A}}}{\eta _{\rm{B}}}{e_0} + {\gamma _{\rm{T}}}/2}}{{{\eta ^\tau }B{\eta _{\rm{A}}}{\eta _{\rm{B}}} + {\gamma _{\rm{T}}}}}
 \end{equation}
 where ${\eta ^\tau }B{\eta _{\rm{A}}}{\eta _{\rm{B}}}{e_0}$ represents the errors arising from the non-ideal characteristics of the devices. The QBER induced by clock uncertainty is ${\gamma _{\rm{T}}}/2$. 
 \begin{equation}
 \gamma  = \left( {B{\eta _A} + 2{\rm{D}}{{\rm{C}}_{\rm{A}}}} \right)\left( {B{\eta _B} + 2{\rm{D}}{{\rm{C}}_{\rm{B}}}} \right)\Delta T
 \end{equation}
 where $B$ represents the brightness of the light source, \( \eta_A \) and \( \eta_B \) denote the detection efficiencies, incorporating losses from the free-space channel, and \( D_A \) and \( D_B \) correspond to the dark count rates of the detectors.
 The asymptotic secret key rate is estimated through $ R = Q_\text{A}(1-(1+f)H_2(E))$, where $Q_\text{A} = Q_\text{z} +Q_\text{x} $ represent the total coincidence count across both measurement bases, $H_2(\cdot)$ denotes the binary entropy function of the QBER, and $f$ characterizes the error correction efficiency.

\begin{algorithm*}
  \caption{Dual-Pointer Fast Correlation Algorithm}\label{alg:dual_pointer}
  \SetAlgoLined
  \KwIn{$\mathcal{T}_A[0..N_A-1]$, $\mathcal{T}_B[0..N_B-1]$: Timestamp sequences \\
        $N$: Number of histogram bins \\
        $\Delta t$: Time bin width \\
        $\Delta_\text{offset}$: System delay compensation}
  \KwOut{$H[0..N-1]$: Normalized intensity correlation histogram}
  
  Initialize $H \leftarrow \mathbf{0}_{N}$ \tcp*{Zero-initialized histogram}
  $i \leftarrow 0$, $j \leftarrow 0$ \tcp*{Pointer initialization}
  $L \leftarrow -\lfloor N/2 \rfloor$, $U \leftarrow N - \lfloor N/2 \rfloor$ \tcp*{Correlation window bounds}
  
  \While{$i < N_A \land j < N_B$}{
      $k \leftarrow \lfloor \mathcal{T}_A[i]/\Delta t \rfloor - \lfloor \mathcal{T}_B[j]/\Delta t \rfloor - \Delta_\text{offset}$ \tcp*{Binning calculation}\label{line:binning}
      
      \eIf{$k < L$}{
          $i \leftarrow i + 1$ \tcp*{Advance $\mathcal{T}_A$ pointer}
      }{
          \eIf{$k \geq U$}{
              $j \leftarrow j + 1$ \tcp*{Advance $\mathcal{T}_B$ pointer}
          }{
              $H[k - L] \leftarrow H[k - L] + 1$ \tcp*{Histogram update}\label{line:hist_update}
              $i \leftarrow i + 1$, $j \leftarrow j + 1$ \tcp*{Synchronized advance}
          }
      }
  }
  \Return{$H$}
  \end{algorithm*}
 
\section{\label{Appendix:G} Fast entangled coincidence measurement correlation algorithm}
The entangled coincidence measurement requires computing intensity correlation functions through convolution operations on high photon count rate. To address the stringent computing resource constraints of drone platforms and meet real-time processing requirements, we developed a dual-pointer fast correlation algorithm achieving linear time complexity $O(n)$, as detailed in Algorithm 1. The algorithm's core innovation lies in its dual-pointer sliding-window mechanism and ordered-data processing, which reduce computational complexity.The timestamps are inherently time-ordered-a property guaranteed by their physical generation process. By synchronizing the traversal of pointers $i$ and $j$, the algorithm requires a single pass through the data.

The algorithm were executed on a consumer-grade mini personal computer, the specification is listed in  as follows Tables~\ref{tab:pc_configurations} , demonstrating the protocol's compatibility with low power computing platforms.  
Despite the modest 1.60 GHz base clock speed and 4-core configuration, our algorithm successfully processed data block containing $144\times10^3$ (Alice) and $133\times10^3$ (Bob) detection events within a 1000 ns relative delay window (20000 bins and bin with of 50 ps). Execution time was measured using std$::$chrono$::$high\_resolution\_clock from the C++ Standard Library, yielding an average correlation computation time of $7.49\pm0.74$ ms over 10 trial.This performance confirms the efficiency of our optimized dual-pointer intensity correlation function and demonstrates its practical deployability in resource-constrained application.
\begin{table*}
\caption{\label{tab:pc_configurations}%
Hardware specifications for real-time quantum synchronization processing.}
\begin{ruledtabular}
\begin{tabular}{ccccc}
\textrm{Component}&
\textrm{Processor}&
\textrm{Operating}& 
\textrm{Development}&
\textrm{Programming}\\   
\colrule
Specification & \makecell{Intel Core i5-10210U \\ (4 cores$@$1.60 GHz)} & \makecell{Windows 10 64-bit \\ (8 GB RAM) } & \makecell{Microsoft Visual Studio \\ Community 2019 \\ (Version  $16.11.44$)}  & C++
\end{tabular}
\end{ruledtabular}  
\end{table*}

\section{\label{Appendix:H} Literature review}
Recent advances in entanglement distribution ,including implementations across both free-space (FS) and fiber channels, reveal three distinct synchronization approaches with characteristic performance profiles (Table~\ref{tab:qkd_comparison}). (i) Standalone clock systems relying on GNSS or Rb-clock references exhibit fundamental precision limitations, where Rb-clocks achieve hundred-picosecond root-mean-square (RMS) jitter \cite{krzicMetropolitanFreespaceQuantum2023} while GNSS typically exceeds 10 ns in coincidence window \cite{PRXQuantum.2.040304}. (ii) Hybrid classical systems combining GNSS with Rb-clocks and synchronization pulses demonstrate 13 ps RMS jitter\cite{neumannContinuousEntanglementDistribution2022}, GNSS With Synchronization pulse enabling operation in challenging environments including satellite-to-ground links \cite{Yin_2017,yinEntanglementbasedSecureQuantum2020} and subsea fiber deployments \cite{wengerowskyEntanglementDistribution96kmlong2019}, and (iii) Entanglement-corrected (EC) hybrid architectures that utilize classical references for initial coarse synchronization followed by entanglement-assisted refinement, 12 ps when integrated with high-stability Rb-clocks.

\begin{table*}
\caption{\label{tab:qkd_comparison}
Literature review.}
\begin{ruledtabular}
\begin{tabular}{ccccccccc}
Reference & \makecell{Time Sync \\ Method} & \makecell{Alice(Bob) \\ Count rate (cps)} & \makecell{Coincidence \\ (cps)} & \makecell{Loss \\ (dB)} & \makecell{RMS jitter \\ (ps)}  & \makecell{$\tau_w$ \\ (ns)} & \makecell{Link type \\ (distance)}  \\
\hline
\makecell{Yin et al. \\ 2017\cite{Yin_2017}} & \makecell{GNSS \& \\ Sync pulses} & - & 37.8 & 29$\sim$44 & 700 & 2 & \makecell{FS (530km \\ $\sim$1600km)}  \\

\makecell{Yin et al. \\ 2020\cite{yinEntanglementbasedSecureQuantum2020}} & \makecell{GNSS \& \\ Sync pulses} & - & 2.2 & 56--71 & 700 & 2.5 & \makecell{FS \\ (1,120 km)} \\

\makecell{Ecker et al. \\ 2021\cite{eckerStrategiesAchievingHigh2021}} & \makecell{GNSS \& \\ EC} & \makecell{13.3$\times10^{3}$ \\ (10$\times10^{3}$)}  & $>$300 & 38.72 & ~ & 1.0 & \makecell{FS \\ (143 km)} \\

\makecell{Basset et al.\\ 2021\cite{doi:10.1126/sciadv.abe6379}} & \makecell{GNSS \& \\ EC} & \makecell{700$\times10^{3}$ \\ (620$\times10^{3}$) }& - & - & - & 1.6 & \makecell{FS \\ (270 m)} \\

\makecell{Basset et al. \\ 2023\cite{bassobassetDaylightEntanglementbasedQuantum2023}} & \makecell{ GNSS \& \\ EC} & \makecell{470$\times10^{3}$ \\ (58$\times10^{3}$)} & 106 & - & - & 1.3 & \makecell{FS \\ (270 m)} \\

\makecell{Bulla et al. \\ 2023\cite{PhysRevX.13.021001}} & \makecell{GNSS \& \\ Rb-clock \& \\  EC} & - & \makecell{140 $\sim$ \\ 505} & 25 & 20 & - & \makecell{FS \\ (10.2 km)}\\

\makecell{Kržič et al. \\ 2023\cite{krzicMetropolitanFreespaceQuantum2023}} & Rb-clock & \makecell{1$\times10^{6}$ \\ (190$\times10^{3}$)} & - & $\sim$0 & - & 0.71 & \makecell{FS \\ (1.7 km)}\\

\makecell{Spiess. \\ 2023\cite{PhysRevApplied.19.054082}} & \makecell{NTP \& \\ EC} & \makecell{(195$\pm$3$)\times10^{3}$ \\ (15$\pm$5$)\times10^{3}$} & 400$\pm$200 & $\sim$0 & $<$68 & 0.71 & \makecell{FS} \\

\makecell{Wengerowsky et al. \\ 2019\cite{wengerowskyEntanglementDistribution96kmlong2019}} & \makecell{GNSS \& \\ Sync pulses} & \makecell{1.3$\times10^{6}$ \\ (1.93$\times10^{6}$)} & 505 & 22 & 400 & $\sim$0.7 & \makecell{Fiber \\ (96 km)}\\

\makecell{Alshowkan et al. \\ 2021\cite{PRXQuantum.2.040304}} & GNSS & - & $>$26 & 1.8$\sim$3.3 & \makecell{$1.21 \times10^4$ \\ $1.47 \times10^4$}  & 10 & \makecell{Fiber \\ (Network)}\\

\makecell{Neumann et al. \\ 2022\cite{neumannContinuousEntanglementDistribution2022}} & \makecell{ GNSS \& \\ Rb-clock} & \makecell{62.5$\times10^{3}$ \\ (94.4$\times10^{3}$)} & 8.9 & 64.5 & 13 & 0.114 & \makecell{Fiber \\ (248 km)} \\

\makecell{Liu et al. \\ 2023\cite{Liu_2024}} & GNSS & \makecell{6.9$\times10^{3}$ \\ (2.7$\times10^{3}$) }& 0.59 & 51 & - & 0.24 & \makecell{Fiber \\ (242 km)}\\

\makecell{Pelet et al. \\ 2025\cite{peletEntanglementbasedClockSyntonization2025}} & \makecell{ Rb-clock \& \\ EC} & \makecell{10$\times10^{6}$ \\ (3$\times10^{6}$)} & - & 13 & $<$12 & 0.12 & \makecell{Fiber \\ (50 km)} \\
\end{tabular}
\end{ruledtabular}
\end{table*}

\nocite{*}

\bibliography{apssamp}

\end{document}